\def\TitleOfPaper{Correlation energy of the uniform gas determined by ground state conditional probability density functional theory}
\def\preprintlinklocation{http://dft.uci.edu} %Need to add paper ref when uploaded

\documentclass[twocolumn, secnumarabic,  floatfix, nofootinbib, nobalancelastpage]{revtex4-2}

\usepackage{amsmath}
\usepackage{physics}
\usepackage{amssymb}
\usepackage{float}
\usepackage[dvipsnames]{xcolor}

\definecolor{TITLECOL}{rgb}{0.05,0.25,0.85}
\definecolor{CONTENTSCOL}{rgb}{0.1,0.2,0.7}
\definecolor{URLCOL}{rgb}{0,0.52,0.83}
\definecolor{LINKCOL}{rgb}{0.05,0.5,0}
\definecolor{CITECOL}{rgb}{0.25,0,0.48}
\definecolor{SECOL}{rgb}{0.07,0.31,0.80}
\definecolor{SSECOL}{rgb}{0.26,0.19,0.75}
\definecolor{SSSECOL}{rgb}{0.26,0.19,0.75}

\newcommand{\coloredtitle}[1]{\title{\textcolor{TITLECOL}{#1}}}
\newcommand{\coloredauthor}[1]{\author{\textcolor{CITECOL}{#1}}} 
\renewcommand{\sec}[1]{\section{\textcolor{SECOL}{#1}}}
\newcommand{\ssec}[1]{\subsection{\textcolor{SSECOL}{#1}}}

\usepackage{graphicx}

\def\PublicationsLink{http://dft.uci.edu/publications.php}

\def\preprinttext{~}

\usepackage{fancyhdr}
\makeatletter
\fancypagestyle{titlepage}
{	\lhead{}
	\chead{}
	\rhead{}
	\lfoot{}}
\makeatother

\definecolor{Green}{rgb}{0.016,0.627,0}
\definecolor{Plum}{rgb}{0.17,0,0.45}
\definecolor{LBlue}{rgb}{0.18,0.59,1}
\definecolor{Sepia}{rgb}{0.37,0.17,0.02}
\definecolor{BurntOrange}{rgb}{0.78,0.39,0}

\usepackage{hyperref}
\hypersetup{
    breaklinks=true,
    bookmarksopen=true,
    bookmarksnumbered=true,
    colorlinks=true,
    linkcolor=LINKCOL,
    linktocpage=true,
    citecolor=CITECOL,
    filecolor=magenta,      
    urlcolor=URLCOL,
}

\def\bea{\begin{eqnarray}}
\def\eea{\end{eqnarray}}
\def\ben{\begin{equation}}
\def\een{\end{equation}}
\def\benu{\begin{enumerate}}
\def\enu{\end{enumerate}}

\def\bei{\begin{itemize}}
\def\eei{\end{itemize}}
\def\beit{\begin{itemize}}
\def\eit{\end{itemize}}
\def\benu{\begin{enumerate}}
\def\enu{\end{enumerate}}

\def\n{n}

\def\sss{\scriptscriptstyle\rm}

\def\1var{(\bx_1...\bx\N)}

\def\br{{\bf r}}

\def\bx{{x}}

\def\x{_{\sss X}}
\def\c{_{\sss C}}
\def\s{_{\sss S}}
\def\xc{_{\sss XC}}

\def\Hxc{_{\sss HXC}}

\def\N{_{\sss N}}
\def\H{_{\sss H}}

\def\sph_int{ {\int d^3 r}}

\usepackage{cases}
\usepackage[normalem]{ulem}
\def\brr{\br}
\def\brp{\br^{\prime}}

\begin{document}
\sf
\coloredtitle{\TitleOfPaper}

%Put the names of authors here:
\coloredauthor{Dennis Perchak}
\email{dperchak@icloud.com }
\affiliation{Department of Chemistry, University of California, Irvine, CA, 92697}

\coloredauthor{Ryan J. McCarty}
\email{rmccarty@uci.edu}
\affiliation{Department of Chemistry, University of California, Irvine, CA, 92697}

\coloredauthor{Kieron Burke}
\email{kieron@uci.edu}
\affiliation{Department of Chemistry, University of California, Irvine, CA, 92697}
\affiliation{Department of Physics and Astronomy, University of California, Irvine, CA, 92697}
\date{\today}

\begin{abstract}
Conditional-probability density functional theory (CP-DFT) is a formally exact
method for finding correlation energies from Kohn-Sham DFT
without evaluating an explicit energy functional.  
We present details on how to generate accurate exchange-correlation 
energies for the ground-state uniform gas.  We also use the exchange
hole in a CP antiparallel spin
calculation to extract the high-density limit.
We give a highly accurate analytic solution
to the Thomas-Fermi  model for this problem,
showing its performance relative to Kohn-Sham and may be useful at high temperatures.
We explore several approximations to the CP potential.
Results are compared to accurate parameterizations for both exchange-correlation energies and holes.
\end{abstract}
\maketitle
%\tableofcontents

\sec{Introduction}

The uniform electron gas (UEG) plays an iconic role in condensed matter physics\cite{GG08}.
It is the simplest example of an interacting electronic system in the thermodynamic limit.
It has also been used as a simple model for metals, especially simple metals like Na and Al\cite{JG89}.
Moreover, its exchange-correlation (XC) energy is a vital input to the local
density approximation (LDA) for inhomogeneous systems, an approximation that dominated
DFT calculations for a generation, and remains in wide use today\cite{B12}.
The ground-state energy was first calculated accurately by Ceperley and Alder\cite{CA80},
and modern parameterizations largely agree at the 1\% level\cite{VWN80,PZ81,PW92}.   On the other hand,
the finite temperature case is still actively being calculated today\cite{KSDT14,GDSMFB17, DCRB20}.

Recently, it was pointed out that, for any electronic system, one could extract
the unknown XC energy by a sequence of Kohn-Sham DFT (KS-DFT) calculations (one for each point
on a grid in the system) if a well-defined but unknown potential, the conditional probability
potential, were known\cite{MPPE20b}.
A simple LDA to this potential
yields surprisingly accurate results for systems as disparate as
the binding energy curve of H$_2$ and the uniform gas.  This approximation
automatically has no self-interaction error for one-electron systems\cite{PZ81}, correctly
dissociates the H$_2$ singlet into two separate H atoms\cite{MCY08}, and its accuracy does
not deteriorate as the temperature is raised in a uniform gas\cite{MPPE20b}.

However, many questions were left unexplored in the original publication.   The
formal underpinnings, for any quantum system, are being presented elsewhere \cite{PWB21}.
In the present work, we focus exclusively on the uniform gas at zero temperature.
We give details on how CP calculations are performed, which differ substantially
from a traditional DFT calculation, both in the nature of the external potential
and the boundary conditions.
We first perform the crudest possible CP calculation,
by adding a simple Coulomb repulsion at
the reference point.
This is called a blue electron approximation, thinking of the reference electron as distinguishable from all others (painted blue). The CP is approximated as the ground-state density of an $N-1$-electron system with the blue electron treated as an external potential generating a simple Coulomb repulsion.This scheme works remarkably well for densities with Wigner-Seitz radii, $r_s > 2$,
with errors of less than 12\% for the XC energy.  However, it fails badly at high densities where exchange dominates, and so is poorly modeled 
by a classical approximation.  Essentially, the on-top exchange hole
and its environs is about minus half the density for an unpolarized gas, while
the blue electron potential digs almost no ontop hole at all.

We next show how this difficulty was overcome in Ref \cite{MPPE20b}.   First, the pure blue
electron approximation was shown to violate the electron-electron cusp condition
by a factor of 2.  When this is restored, accurate results for highly quantum
systems were achieved, but it has little or no impact on the high-density uniform
gas.   Then, a large repulsive Gaussian was added to the CP potential, with
parameters chosen to reproduce the exchange hole, i.e., essentially an accurate
CP potential in this limit.   As this does not require any many-body calculation,
this is deemed acceptable.   But it was also necessary to smoothly turn this
Gaussian off as a function of increasing $r_s$, a somewhat empirical procedure,
but one which yields the high accuracy presented in Ref \cite{MPPE20b}.

Here, we bypass this messy procedure, but still without using any many-body
results for the uniform gas.
We present a more sophisticated and satisfactory solution, by performing a
CP calculation for antiparallel spin only, fixing the parallel CP density to that
of exchange (which has long been known analytically\cite{D30}). By construction, this recovers the correct exchange in the high-density
limit, and yields reasonable accuracy for all other $r_s$, but is less accurate
in the low density limit.  Combining this procedure at high densities with the
original calculation at low densities yields an accurate curve for all $r_s$ values.

Lastly, we consider the Thomas-Fermi (TF) solution to the blue electron problem, and evaluate its accuracy relative to a full KS calculation.
We solve the TF equation analytically using
a high-density approximation, which turns out to be remarkably accurate for all densities.
For atoms, inaccurate TF densities are the chief source of error in TF energies.
But here we are calculating a simple impurity in an otherwise uniform background,
thereby avoiding evanescent regions, etc.
Moreover, since the XC energy involves a double spatial integral in general
(and a single integral), it may be more forgiving of errors than an
explicit density functional might be.
We find that TF CP densities
yield reasonably accurate XC energies in CP theory, and their accuracy
improves with increasing $r_s$.
We expect TF-CP calculations to be particularly useful at temperatures
beyond which the KS equations fail to converge, a regime which inspired many of
these ideas (see Ref.~\cite{C91} and subsequent work).

\sec{Theory}
\subsection{Background}

Standard KS-DFT calculations solve the KS equations within some approximation for the 
XC energy as a functional of the density, $E\xc[\n]$:
\ben
\label{eq_KS}
\left[ -\frac{1}{2} \nabla^2 + v\s[n](\br)\right] \phi_{i}(\br) = \varepsilon_i\, \phi_i(\br),
\een
where $\phi_i$ are the KS orbitals and $\varepsilon_i$ the eigenvalues. Hartree atomic units are used throughout. The KS potential is
given by
\ben
\label{eq_KS_potential}
v_{\sss S}[n](\br) = v(\br) + v_{\Hxc}[n](\br),
\een
where $v(\br)$ is the original one-body potential while $v\Hxc(\br)$ 
includes the Hartree potential and
an XC contribution, $v\xc(\br) =\delta\,E\xc[n]/\delta\,n(\br)$.  In practice, spin-DFT is used.
Most calculations of total energies yield densities sufficiently close to the exact
density that the error in the total energy is dominated by the error in $E\xc$ itself, 
rather than on its evaluation on the approximate density\cite{KSB13}.   
While there have been many improvements and refinements in such approximations over the last half century,
there are many known systematic limitations of present-day functionals, such as their
inability to correctly dissociate bonds (strong correlation effects)
and electron self-interaction errors\cite{KK20}.

Recently, CP-DFT, an alternative approach to electronic structure calculation, was suggested \cite{MPPE20b}.
This takes advantage of the well-known expression for the
XC energy in terms of the (coupling-constant averaged) XC hole, which in turn is simply related to
the conditional probability density for finding an electron at $\br'$, given an electron at $\br$.
Knowledge of this function (or in fact just some particular integrals over it) determines $E\xc$ 
exactly.   The aim in CP-DFT is to \textit{calculate} this CP probability density using a standard
KS calculation at every point $\brr$ in the system.  In principle, if it exists, there is a unique correction
to the one-body potential such that the ground-state of $N-1$ electrons yields the desired CP density~\cite{HK64}. 
In practice, we find a simple local density approximation works very well most of the time, as our results show.

Thus, writing
\ben
E_{\xc} = \frac{1}{2}\int_{0}^{1} d\lambda \int\,d^{3}r\int\,d^{3}r' \; \frac{n(\brr) n^{\lambda}_{\xc}(\brr,\brp)}{|\brr - \brp|},
\label{eq_EXC_pair}
\een
where integrating over $\lambda$ is the adiabatic connection between the KS system at $\lambda = 0$ and interacting system at $\lambda = 1$~\cite{LP75}.
The XC hole density is determined via
\ben
n_{\xc}^\lambda(\brr,\brp) = \tilde{n}^\lambda_\br(\brp)-n(\brp),
\label{eq_XC_density}
\een
where $\tilde{n}^\lambda_\br(\brp)$ is the conditional probability density for finding an electron at $\brp$, given an electron at $\br$, at coupling strength $\lambda$.
We define the CP potential as the potential that
has ground state density $\tilde{n}_{\brr}^\lambda(\brp)$ for $N-1$ electrons. It can be written as $v[\tilde{n}_{\brr}^\lambda](\brp)$, where $v[n](\brr)$ is the functional dependence of the potential on the ground state density, first described by Hohenberg and Kohn~\cite{HK64}. We then define:
\ben
\label{eq_CorPot}
\Delta \tilde{v}_{\brr}^\lambda (\brp) = v[\tilde{n}_{\brr}^\lambda](\brp) - v^\lambda(\brp)
\een
as the correction to the external potential needed to yield the CP potential. 
This is a functional of the original ($N$-electron) gas density $n(\brr)$, by the Hohenberg-Kohn theorem\cite{HK64}.

This theory is formally exact if the potential exists, but the exact CP potential is not known in general and must be approximated.
A simple approximation, $\Delta \tilde{v}_{\br}^\lambda = \lambda/|\brr-\brp|$, accounts for the majority of the CP correlation, and becomes 
exact as the distance between $\brr$ and $\brp$ increases. 
This can be considered the correct potential in the classical limit, in which particles can be distinguished
from one another, and where the missing electron provides a simple Coulomb repulsive impurity potential.

But this cannot be exact for quantum systems in general.  For example,
recovery of the electron-electron cusp conditions at small $|\brr - \brp|$ distances requires instead $\lambda/(2|\brr-\brp|)$. 
A linear interpolation between small and large distances, depending locally on the density, yields
\ben
\label{eq_vcprl_Erf}
\Delta \tilde{v}_{ \brr}^\lambda[n](\brp)\approx\frac{\lambda}{2|\brr-\brp|} \left(1+{\rm\,Erf}\left(\frac{|\brr-\brp|}{r_s(\n(\brr))}\right) \right),
\een
where $r_s=(3/(4\pi\n))^{1/3}$ is the Wigner-Seitz radius of the gas density at the reference point, provides a simple approximation that is correct in both limits \cite{MPPE20b}.
While this mirrors the longstanding challenge of approximating 
$E_{\xc}$ with a density functional, 
DFT methods often produce highly accurate densities even when
their energies are incorrect \cite{KSB13}, allowing CP-DFT to produce accurate solutions for several systems that are challenging for standard DFT methods.

For those familiar with classical DFT of distinguishable particles,
this procedure is very like the 
Percus-Yevick closure\cite{PY58} of the Ornstein-Zernike equation\cite{OZ14}.
In the truly classical case, the blue electron approximation would be exact,
if the exact inhomogeneous XC functional were used when solving the KS equations
for the impurity.   In practice, we expect the use of approximate XC in 
such calculations to produce only very slight errors, much of which will be forgiven by
the multiple integrals in Eq.~\eqref{eq_EXC_pair}.   
Moreover, we anticipate that errors will reduce as the density becomes lower, the
electrons become more spread out, and the behavior is dominated by pure Coulomb repulsion.
Likewise, in the high density limit, we expect difficulties, where the entirely non-classical
exchange effect dominates.

This paper reports only results for the uniform gas at zero temperature.   It gives the full
details of the potentials and computational methods that were used in \cite{MPPE20b}, plus an
extension of the method used there to extract spin-decomposed holes, and better justify
the smooth turn-off of the exchange barrier needed to recover the high-density limit.
We also discuss the quasi-analytic solution of the Thomas-Fermi equation (instead of the
KS equations) for this case.   This is of interest, as clearly a TF solution is far less
computationally expensive than the KS approach.   In cases where the additional cost of 
solving KS equations at every point is prohibitive, use of TF theory may provide
a practical alternative, but only if errors are not substantially increased.

\sec{Methods}

\ssec{CP-DFT}

We now consider applying the CP concept to the spin unpolarized uniform electron gas.  By symmetry, the CP potential is spherical, and we perform calculations within a finite sphere of radius $R$.
Because the UEG is translationally invariant, we choose $r=0$ for the reference point. Further, we drop the prime from $r^\prime$, so that $r$ now denotes the distance from the reference point, as is conventional for UEG calculations.  In practice, we perform calculations in a finite sphere with a large but finite number of electrons $N$.  The boundary of the sphere produces non-uniformities in the density, which are minimized by a judicious choice of boundary condition, Eq.~\eqref{eq:noflux2}.  Moreover, by calculating our XC hole density, Eq.~\eqref{eq:hole}, by subtracting densities with $N$ and $N-1$ electrons, the effect of non-uniformity near the surface largely cancels.  Thus all our calculations are converged with respect to the size of the sphere.
The systems have a maximal radial coordinate $R$, volume of $V = 4\pi R^3/3$, and average electron density of $\bar{n} = N/V$ and $(N-1)/V$. The Kohn-Sham (KS) equations are solved in the local density approximation (LDA) for both systems.
The potential used in the KS equations is
\begin{equation}
\label{KSBlueNS}
v\s[n](\br) = v_{\sss n}[n_0](\br) +  \Delta \tilde{v}[n](\br) + v\H[n](\br) + v\xc[n](\br),
\end{equation}
where $v\H[n](\br)$ is a Hartree repulsion 
\begin{equation}
\label{ve}
v_H(r) = \int d^3r' \frac{n(\mathbf{r'})}{|\mathbf{r-r'}|},
\end{equation}
and $v_{\sss n}[n_0](\br) $ is a potential derived from a uniform, positive compensating background charge density, $n_0 = N/V$.
The exchange-correlation potential, $v\xc[n](\br)$ is from LDA using the PW92 parametrization \cite{PW92}.
Lastly, we have the CP or blue potential, which appears only in the $N-1$ electron calculation for which we will consider two forms,
\begin{subnumcases}{\label{vBKS} \Delta \tilde{v}(r^\prime)=}
\frac{1}{r^\prime} \label{simple} \\
\frac{1}{2 r^\prime} \left( 1+{\rm\,Erf}\left(\frac{r^\prime}{\bar{r}_s}\right)  \right). \label{system}
\end{subnumcases}
Here, $\bar{r}_s$ refers to the Wigner-Seitz radius of the system as a whole. As was discussed in the introduction, the simple approximation of $1/r^\prime$ for the CP potential is not sufficient for quantum systems. However, it is still instructive  to see where it fails.
Further details of the CP-DFT and the numerical solution method are contained in Appendix A.

\ssec{CP-DFT results}

\begin{figure}[htb]
\begin{center}
\includegraphics[angle=0,width=8.5cm]{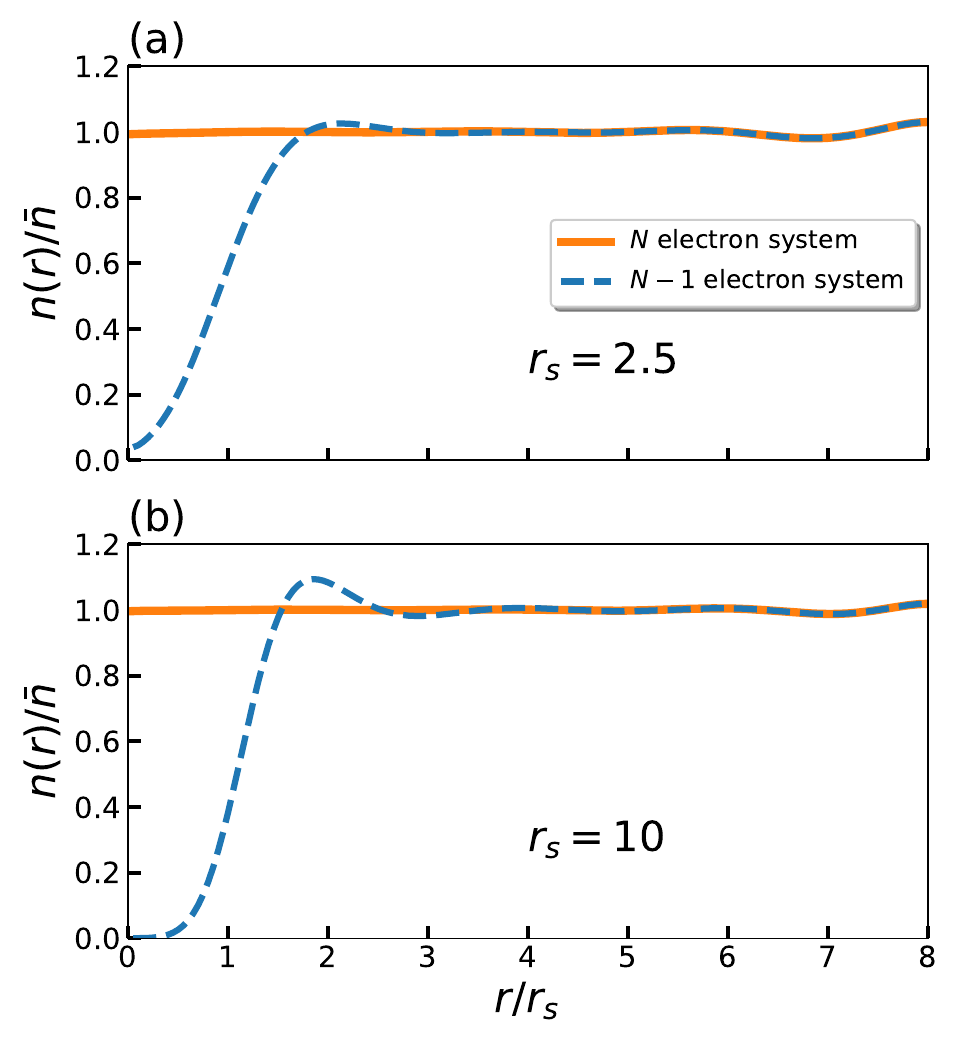}
\end{center}
\caption{Normalized  density vs $r/r_s$ for (a) $r_s = 2.5$ and (b) $r_s = 10.0$. Solid line is the $N$-electron system and the dashed line is the blue electron, $N-1$ system.}
\label{f:f_1}
\end{figure}

\begin{figure}[htb]
\begin{center}
\includegraphics[angle=0,width=8.5cm]{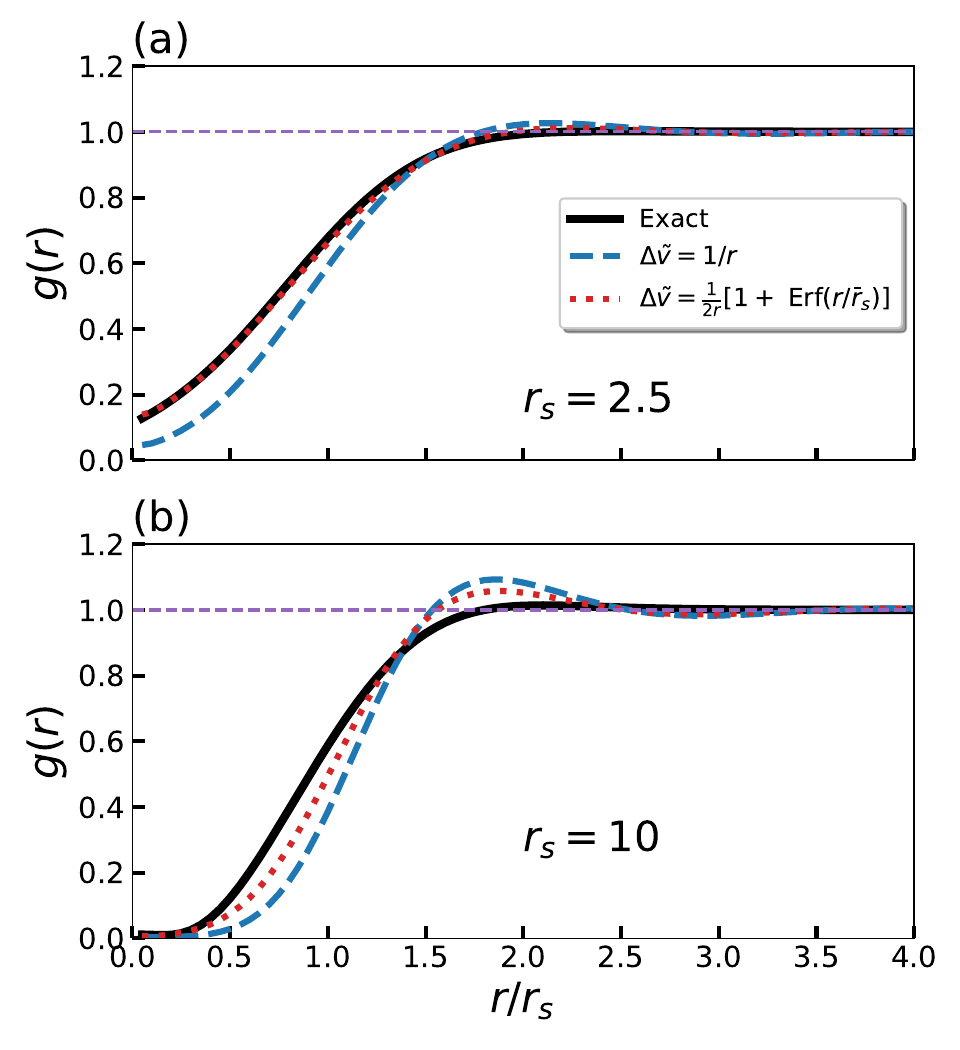}
\end{center}
\caption{Pair distribution function, $g(r)$ plotted vs $r/r_s$ for (a) $r_s = 2.5$ and (b) $r_s = 10.0$. The dashed blue line is for the CP potential equal to $1/r$, the dotted red line is for the modified CP Potential and the solid black line is the exact solution.}
\label{f:f_2}
\end{figure}

\begin{figure}[htb]
\begin{center}
\includegraphics[angle=0,width=8.5cm]{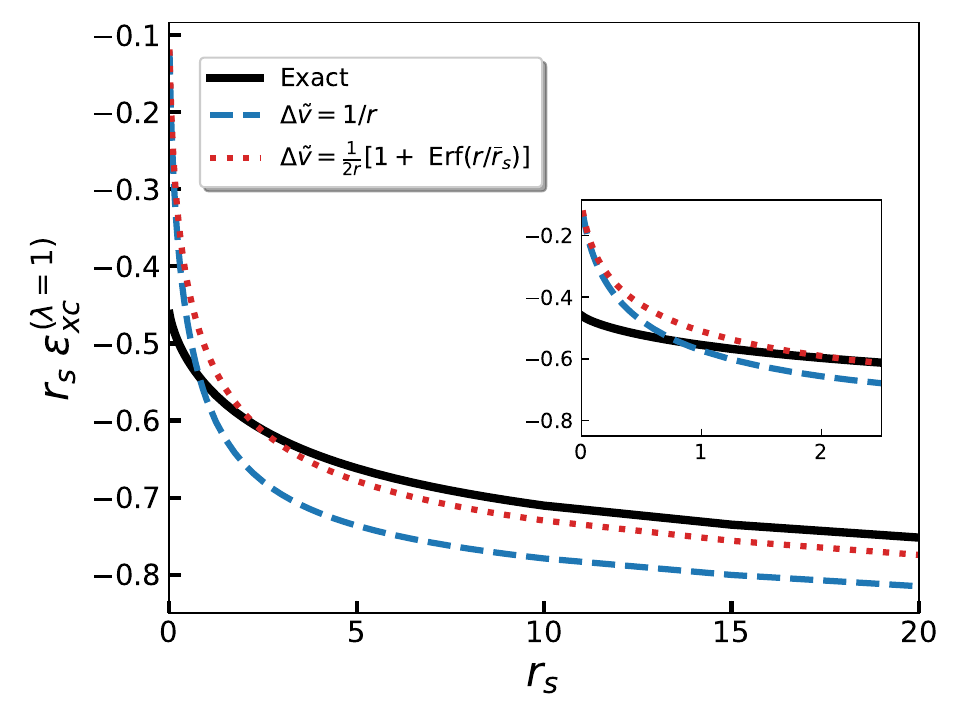}
\end{center}
\caption{$r_s \varepsilon_{xc}^{(\lambda=1)}$ vs $r_s$ for the CP potential equal to $1/r$ (dashed blue line), modified CP potential (dotted red line) and the exact solution (solid black line). The inset shows an expanded view of the region for $r_s \lessapprox 2.0$}
\label{f:f_3}
\end{figure}

\begin{figure}[htb]
\begin{center}
\includegraphics[angle=0,width=8.5cm]{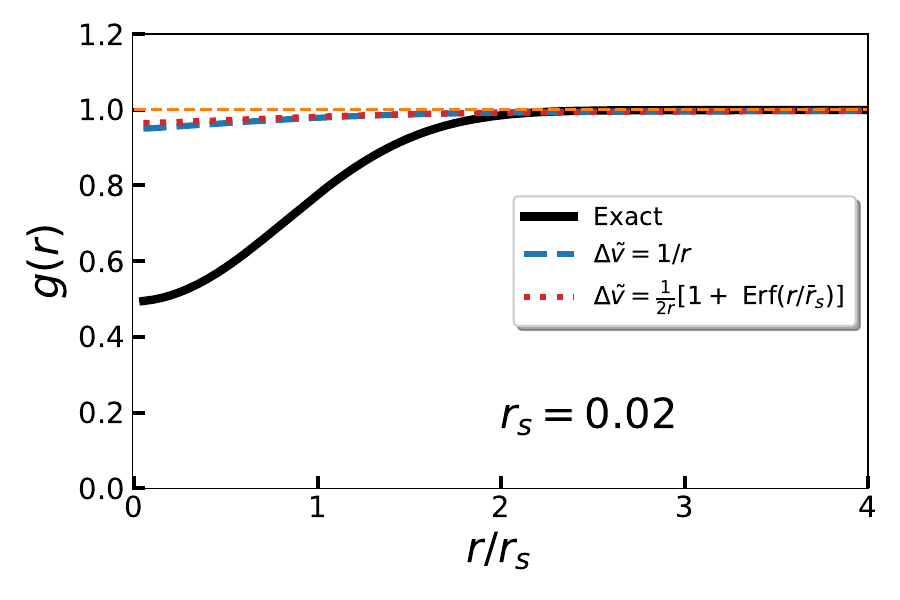}
\end{center}
\caption{Pair distribution function, $g(r)$ plotted vs $r/r_s$ for $r_s = 0.02$. The dashed blue line is for the CP potential equal to $1/r$, the dotted red line is for the modified CP Potential, and the solid black line is the exact soultion.}
\label{f:f_4}
\end{figure}

We first show results for calculations at $r_s =2.5$ and $r_s = 10.0$ using the simple $1/r$ potential, Eq.~\eqref{simple}, for the blue electron.
Fig.~\ref{f:f_1} plots the density of the $N$ electron system and the $N-1$ electron system as a function of the radial coordinate for these two values of $r_s$. 
The blue system density in both cases is depressed near $r = 0$ by the repulsion as we would expect.  The density difference is the XC hole (at $\lambda=1$) from which one deduces the pair distribution function (PDF), with Eq.~\eqref{eq:hole} and Eq.~\eqref{eq:pdf}.
This is shown in Fig.~\ref{f:f_2} where we have also included calculations for the modified CP potential Eq.~\eqref{system}.The advantage of calculating both the $N$ and $N-1$ electron systems is apparent in the cancellation of  boundary effects. The modified CP potential is significantly better than the simple $1/r$ potential.
We can use Eq.~\eqref{eq:exc} to evaluate the XC energy. This energy is \textit{not} averaged over the coupling constant, i.e., it is the potential contribution only. Fig.~\ref{f:f_3} shows $r_s$ times the potential XC energy $\varepsilon_{xc}^{(\lambda=1)}$ plotted vs $r_s$ for the same blue potentials as before, along with the exact values.  The latter is calculated using the PW92 expressions for $g(r)$ along with Eq.~\eqref{eq:pdf} and Eq.~\eqref{eq:exc}.
For values of $r_s \gtrapprox 2.5$, the modified blue potential does much better than the simple Coulomb potential, with relative errors less than 3\% and 11\%, respectively. However, both do badly as the system density increases ($r_s \rightarrow 0$), as shown for $r_s = 0.02$ in Fig.~\ref{f:f_4}.  The model fails to adequately depress the density near $r = 0$. This is due to the dominance of exchange at high density, which is not included in our approximate CP potential. In practical terms, the blue electron is not "repulsive" enough.  A  more satisfactory solution can be obtained by performing a CP calculation in one spin channel only where the other spin CP density is fixed to that given by exchange alone. We turn to the details of this approach in the next section.

\ssec{Spin CP-DFT}

In this section, we perform a spin CP-DFT calculation instead of a CP-DFT
calculation, showing that the high-density limit can be treated accurately
by this method.

The spin conditional probability densities for a system are defined as
the probability density for finding an electron at $\brp$, given an
electron {\em of spin} $\sigma$ at $\brr$. 
Spin conditional probability densities are total densities (not spin densities), but are conditional on an electron having a given spin as well as given position \cite{PWB21}.
We can further decompose
a spin CP density into a sum of two spin densities.  Choosing the blue
electron to have spin up (arbitrarily), we write
\ben
\tilde n_{\br\uparrow}(\brp)=\tilde n_{\br\uparrow}(\brp\uparrow)
+\tilde n_{\br\uparrow}(\brp\downarrow).
\een
The first is the parallel spin CP spin density, the latter being
the anti-parallel.   We perform a spin KS-DFT calculation in which
the former is fixed at its exchange form, which is well-known for a 
uniform gas, given explicitly in (\ref{gx}).
In principle \cite{PWB21}, the CP potential should now be two different potentials, one for
each spin channel, and the functional dependence could differ for parallel
and antiparallel spins.  Here we simply use the same CP potential we have
used for non-spin CP-DFT, (Eq.~\eqref{eq_vcprl_Erf}). For finite systems, using spin densities in CP-DFT is complicated, as in general they do not integrate up to integer particle numbers \cite{PWB21}.   Here, this is not a concern, as the particle number is infinite.  Moreover, in our calculation, we use the exchange pair correlation function, which does yield an integer particle number.

We consider a UEG with spin densities $\bar{n}_\uparrow$  and $\bar{n}_\downarrow$. Placing a \textit{blue} $\uparrow$ electron at the origin, $r=0$ and dropping the prime, it will be surrounded by its hole of $\uparrow$ electrons given by:
\begin{equation}
\delta n_\uparrow (r) =  \bar{n}_\uparrow \left[g_{\uparrow \uparrow} (\bar{n}_\uparrow, \bar{n}_\downarrow;r) - 1\right].
\end{equation}
Here, $g_{\uparrow \uparrow} (\bar{n}_\uparrow, \bar{n}_\downarrow;r)$ is the PDF for like-spin electrons separated by a distance $r$ in a uniform gas with given spin densities, $\bar{n}_\uparrow$ and $\bar{n}_\downarrow$. The up-spin density is then given by:
\begin{equation}
\label{eq:upspin}
\tilde n_{\uparrow}( r \uparrow) = \bar{n}_\uparrow + \delta n_\uparrow(r) = \bar{n}_\uparrow g_{\uparrow \uparrow} (\bar{n}_\uparrow, \bar{n}_\downarrow;r)
\end{equation}
With a known $\tilde n_{\uparrow}( r \uparrow)$, we can write down a KS equation for $\tilde n_{\uparrow}( r \downarrow)$ and solve for it within the local spin density approximation. Thus, \textit{given} $g_{\uparrow\uparrow}$, we can calculate $g_{\uparrow\downarrow}$ via CP-DFT. As $g_{\uparrow\downarrow}$ contains no exchange, we can hope to be more accurate in the high density limit.

The numerical solution for the fixed spin case is carried out in a manner analogous to that for 
CP-DFT. Again, we consider two systems. First,  the $N$ electron system with numbers of up and down spins, $N_\sigma = N/2$ and average spin densities, $\bar{n}_\sigma = N_\sigma/V$ and secondly, the $N-1$ electron system with  $N_\uparrow= N/2 - 1$ up spins and $N_\downarrow = N/2$ down spins.
The $N$ electron system is initialized as $n(r \uparrow) = \bar{n}_{\uparrow}$, $n(r \downarrow) = \bar{n}_\downarrow$
and the $N-1$ electron system is initialized as $\tilde{n}_{\uparrow}(r \uparrow) = \bar{n}_{\uparrow}g_{\sss X}(\zeta,k_Fr)$ and $\tilde{n}_\uparrow(r \downarrow) = \bar{n}_{\downarrow}$. Here, $\bar{n}_{\uparrow (\downarrow)}$ is understood as referencing the $N$ or $N-1$ electron system accordingly.
Further details are contained in Appendix B. By using only $g_x$, which can be derived analytically, we avoid using any many-body results. For high densities, this will be extremely accurate.

\ssec{Computational details}
Calculations can be carried out on personal computers, typically lasting minutes or to hours. 
Our model was written in Python 3 utilizing NumPy \cite{Numpy2011}, SciPy \cite{SciPy2019}, Numba \cite{Numba2015}, and an exchange-correlation module “excor.py” written by Kristjan Haule (Physics Dept. Rutgers University).

\ssec{Spin CP-DFT results}

\begin{figure}[htb]
\begin{center}
\includegraphics[angle=0,width=8.5cm]{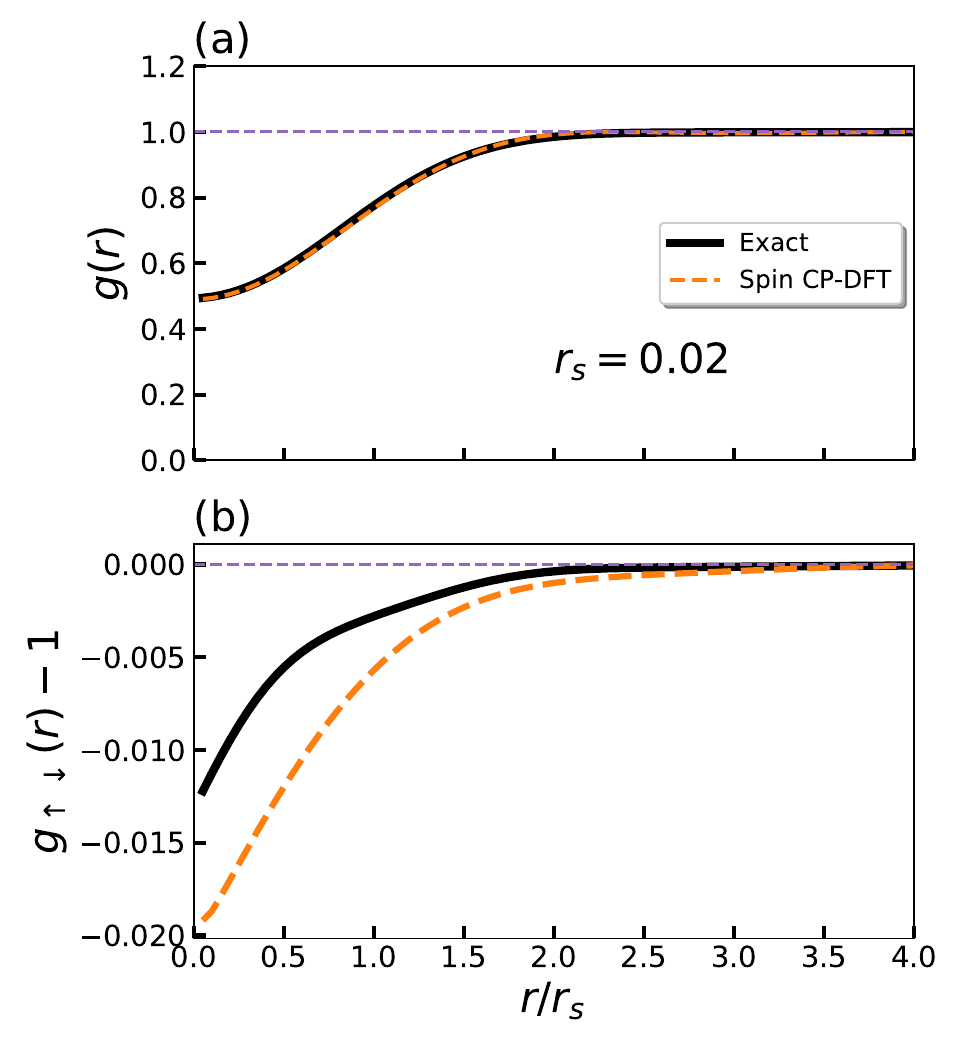}
\end{center}
\caption{Pair distribution function (a) and antiparallel PDF-1 (b) vs $r/r_s$ for the spin CP-DFT with $r_s = 0.02$. Solid black line is the exact result and the orange dotted line is the spin CP-DFT. }
\label{f:f_5}
\end{figure}

The calculations for the spin CP-DFT use the modified version of the CP potential Eq.~\eqref{system}.  Additionally we are setting the correlation potential to zero. We first consider an electron density of $r_s = 0.02$. We plot the result for the PDF in Fig.~\ref{f:f_5}. The agreement is excellent as at such a high density the interactions are dominated by exchange. We also plot $g_{\uparrow \downarrow}(r)$ in Fig.~\ref{f:f_5} and these results are also quite good. 

\begin{figure}[htb]
\begin{center}
\includegraphics[angle=0,width=8.5cm]{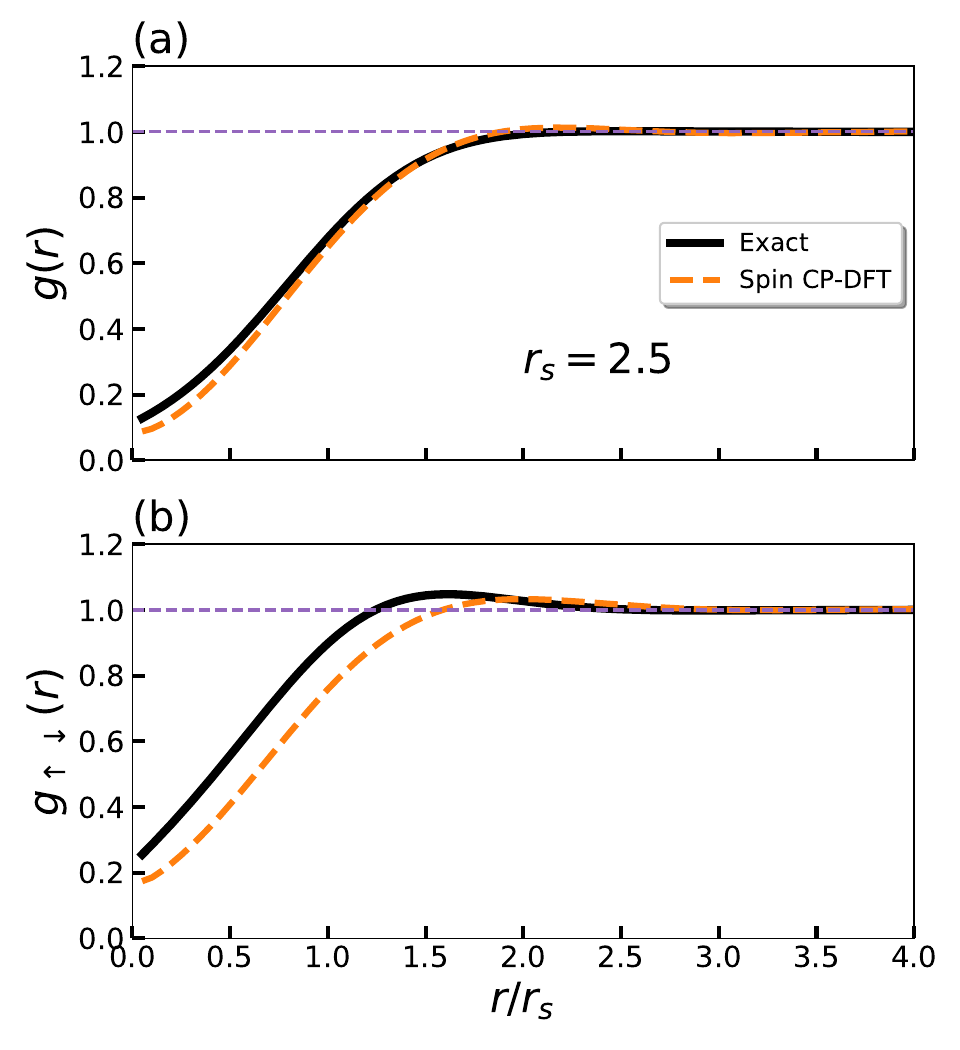}
\end{center}
\caption{Same as for Fig.~\ref{f:f_5} but for $r_s = 2.5$}
\label{f:f_6}
\end{figure}

We next consider a lower density of $r_s = 2.5$.  We show results for the total PDF in Fig.~\ref{f:f_6}(a) and  for the antiparallel PDF in Fig.~\ref{f:f_6}(b). Although there is some small difference with the exact result  for the antiparallel PDF, the overall PDF is good and the error in potential XC energy is about 0.2\%.

\begin{figure}[htb]
\begin{center}
\includegraphics[angle=0,width=8.5cm]{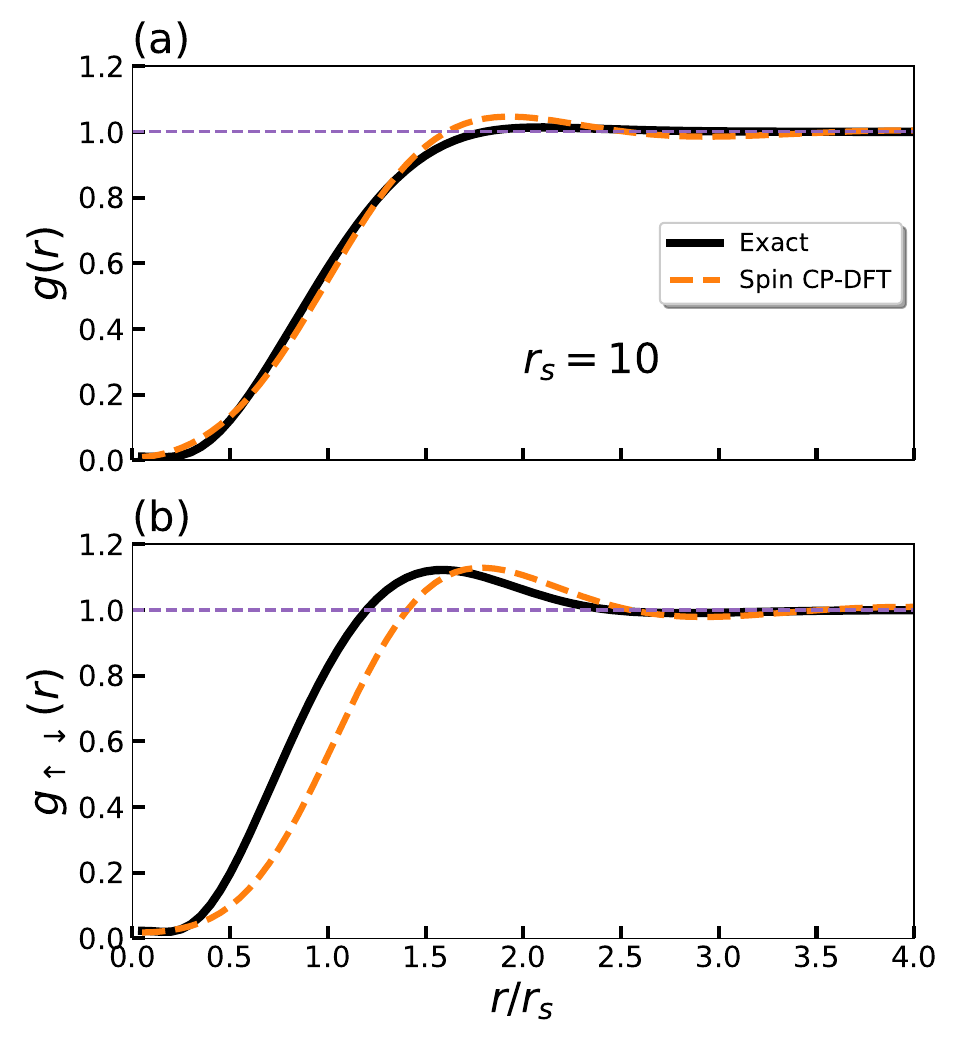}
\end{center}
\caption{Same as for Fig.~\ref{f:f_5} but for $r_s = 10.0$}
\label{f:f_7}
\end{figure}

We see a more serious departure from the exact result at $r_s = 10.0$. Figs.~\ref{f:f_7}(a) and \ref{f:f_7}(b) plot the PDF and antiparallel PDF for this low density. Here we can see more clearly the departure from the exact result and the error in the potential XC energy is about 6\%.

\begin{figure}[htb]
\begin{center}
\includegraphics[angle=0,width=8.5cm]{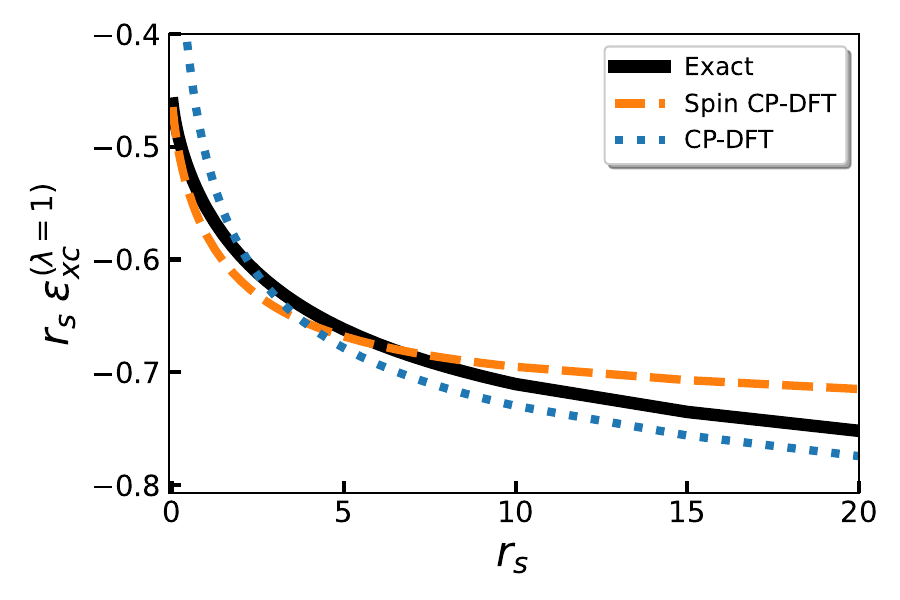}
\end{center}
\caption{$r_s \varepsilon_{\xc}^{(\lambda=1)}$ vs $r_s$. The solid black line is the exact result, the dotted orange line is the spin CP-DFT, and the dotted blue line is for the CP-DFT.}
\label{f:f_8}
\end{figure}

In Fig.\ref{f:f_8} we plot the potential XC energy times $r_s$ as a function of $r_s$ for both models. The agreement with the exact result is quite good for the spin CP-DFT, being worst at low density.  

\begin{figure}[htb]
\begin{center}
\includegraphics[angle=0,width=8.5cm]{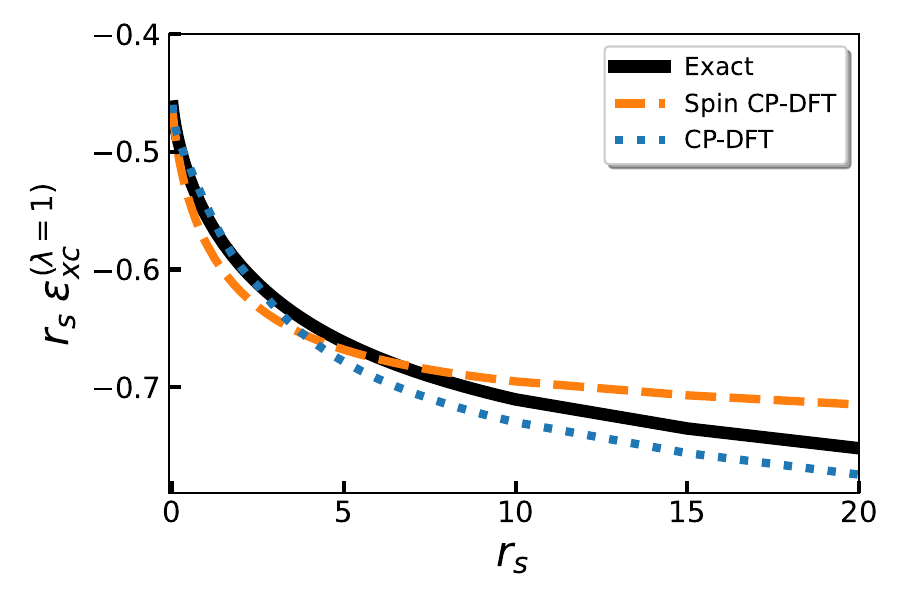}
\end{center}
\caption{Same as for Fig.~\ref{f:f_8} but where the CP-DFT has the added Gaussian potential. }
\label{f:f_9}
\end{figure}

Since, as noted above, for values of $r_s \gtrapprox 2.5$, the fractional error for the CP-DFT is on the order of 3\%, the question arises as to whether an additional potential can be added to the CP-DFT to correct for its deficiencies at high density. In fact, such an additional potential was used in our previous work \cite{MPPE20b}, where we found that an added, external potential of Gaussian shape
\begin{equation}
\label{eq:Gaussian}
v_G(r) = A(r_s) e^{-r^2/2\sigma^2(r_s)},
\end{equation}
where $A(r_s)$ and $\sigma(r_s)$ are $r_s$-dependent strength and range parameters provides the needed repulsive effect.  The $r_s$-dependence was chosen to replicate exchange in the high density limit, and to slowly turn off as $r_s$ grows, vanishing beyond $r_s = 2.5$. This is described in Appendix C. The results of using the Gaussian potential, Eq.~\eqref{eq:Gaussian} with the strength and range parameters given by Eq.~\eqref{eq:powerS} are shown in Fig.~\ref{f:f_9} where the maximum difference with the exact result for the 2-channel model is now less than 3\%. This was used in Ref \cite{MPPE20b}.

\begin{figure}[htb]
\begin{center}
\includegraphics[angle=0,width=8.5cm]{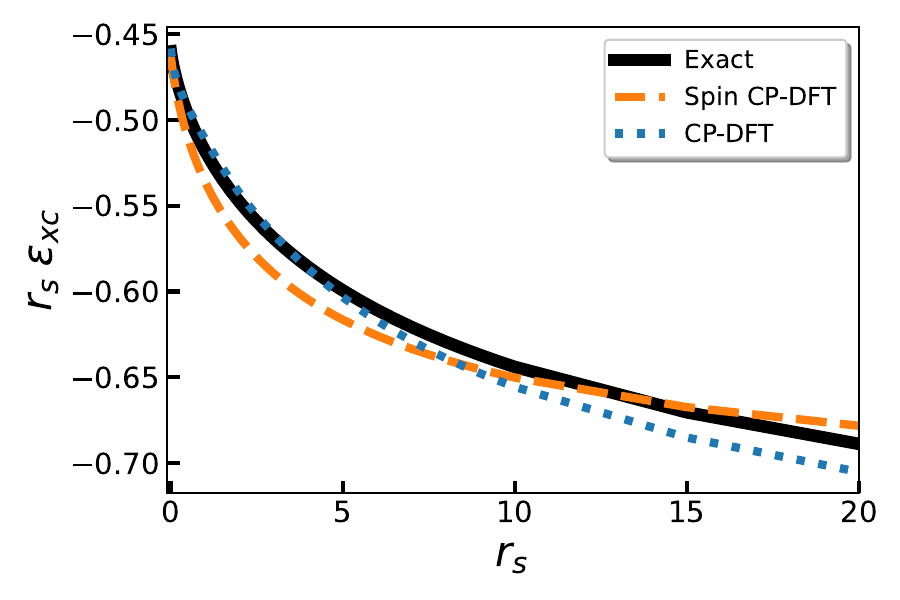}
\end{center}
\caption{Coupling-constant averaged exchange-correlation energies plotted as $r_s \varepsilon_{xc}$ vs $r_s$. The solid black line is the exact result, the dashed orange line is for the spin CP-DFT, and the dotted blue line is for the CP-DFT with the added Gaussian potential. }
\label{f:f_10}
\end{figure}

We can obtain the coupling-constant averaged values of the  XC energy, $\varepsilon_{xc}$, by an integral over $r_s$, i.e.,
\begin{equation}
\label{eq:excBar}
\varepsilon_{\xc}(r_s)  = \frac{1}{r_s^2} \int_0^{r_s}  r_s^{\prime} \varepsilon_{xc}(r_s^{\prime}) dr_s^{\prime}.
\end{equation}
The comparison with PW92 for the XC energy is shown in Fig.~\ref{f:f_10} where the relative difference is below 6\% for the single channel model and for the 2-channel model with Gaussian repulsion is below 3\%.

\sec{Thomas-Fermi Solution for the Blue Electron Problem}

The Thomas-Fermi functional\cite{T27,F28} for the blue electron system  is
\begin{multline}
\label{TFB}
F^{TF}[n] = \int d^3r \left[ A_s n^{5/3}(\mathbf{r}) + v_{\sss n}(\mathbf{r})n(\mathbf{r}) \right. \\
\left. + \frac{1}{2} v_{\sss H}(\mathbf{r})n (\mathbf{r}) + \Delta \tilde{v}(\mathbf{r})n(\mathbf{r}) \right]
\end{multline}
with $A_s =(3/10) \left( 3 \pi^2 \right)^{2/3}$.
The potentials, $v_{\sss n}$, Eq.~\eqref{vn} and $v_{\sss H}$, Eq.~\eqref{ve} are as before and we use $\Delta \tilde{v}(r) = 1/r$.
The charge densities are normalized such that $\int d^3r \; n_0 = N$ and $\int d^3 r \; n(\mathbf{r}) = N-1$.

Defining the total electrostatic potential as $V(\br) = v_{\sss n}(\br) + v_{\sss H} (\br) + \Delta \tilde{v}(\br)$
minimizing the functional, Eq.~\eqref{TFB} with respect to the density, $n(\mathbf{r})$ subject to the normalization condition, and formally solving for the density, we have (just as for bare atoms or ions):
\begin{equation}
\label{nfs}
n(\mathbf{r}) = 
\left( \frac{5}{3} A_s \right)^{-3/2} \left[ \mu - V(\mathbf{r}) \right]_+^{3/2} ,
\end{equation}
where $x_+ = 0$ if $x<0$ and $\mu$ is the chemical potential.

As $r \rightarrow \infty$, $n(r) \rightarrow n_0$, and thus  $V(r) \rightarrow 0$. The latter can be determined from a Taylor expansion of the denominator in $V(\br)$.
From Eq.~\eqref{nfs}, we find
\begin{equation}
\label{Mu}
\mu = \frac{5}{3} A_s n_0^{2/3}.
\end{equation}
But as $r \rightarrow 0$, $V \sim 1/r$, so $\mu - V \rightarrow -\infty$. Thus, $\mu=V$ at some $r_0$, so $n = 0$ for $r \le r_0$ in Eq.~\eqref{nfs}.

We  relate the  Laplacian of the electrostatic potential to the charge densities using Poisson's equation
\begin{equation}
\label{PE0}
\nabla^2 V((\mathbf{r}) ) = -4\pi \left[ n(\mathbf{r}) - n_0 + \delta(\mathbf{r}) \right].
\end{equation}
Defining
\begin{equation}
\label{phiDef}
\frac{\phi(r)}{r} = \mu - V(r)
\end{equation}
and $\delta \phi (r) = \phi (r) - \mu r = -r V(r)$
and substituting  from Eq.~\eqref{nfs} we have
\begin{equation}
\label{PE2}
\nabla^2 \left( \frac{\delta\phi}{r} \right) = 
4\pi \left[-n_0 +  \left( \frac{5}{3} A_s \right)^{-3/2} \left( \frac{\delta\phi}{r} \right)^{3/2} \Theta(r-r_0)  \right], 
\end{equation}
where $\Theta(x)$ is a step function in $x$,
and the $\delta-$function at $r=0$ implies:
\begin{align}
\delta \phi(0) & =  -1, &  \delta \phi(r_0) & =  -\mu r_0. 
\end{align}
Introducing  dimensionless variables $z   = r/r_s$ and $y  = \delta \phi / \mu r_s$ we can write Eq.~\eqref{PE2} as
\begin{equation}
\label{diffY}
\frac{d^2y}{dz^2} = x_s z \left( -1 + \left(1 + \frac{y}{z} \right)_+^{3/2} \right),
\end{equation}
with  $x_s = r_s/a$, and $a = (3\pi^2/16)^{1/3} /2 \approx .61$.
The boundary conditions are:
\begin{align}
\label{BoundY}
y(0) =&  - \frac{x_s}{3}, & y(z_0) =  -z_0, && y \rightarrow 0 \text{ as } z \rightarrow \infty .
\end{align}
For $z < z_0$, 
\begin{equation}
\label{Analyt}
y(z) = - \left( \frac{x_s}{2} \right) z^3 + A z + B,
\end{equation}
where
\begin{align}
\label{BoundYp}
A =&  \left( \frac{x_s}{3z_0} \right) \left[ 1 + \frac{z_0^3}{2} \right] - 1, & B =& -\frac{x_s}{3}.
\end{align}
For $z > z_0$ , we have, for large $z$, to first order
\begin{equation}
\frac{d^2y}{dz^2}  \approx 3x_s \frac{y}{2},
\end{equation}
yielding $y(z) \propto e^{-kz}$ where $k = (3 x_s/2)^{1/2}$.
By matching the log derivatives at $z = z_0$ between Eq.~\eqref{Analyt} and the large $z$ solution, at $z = z_0$, we find
\begin{equation}
\label{cubic}
z_0^3 + \frac{3 k}{x_s} z_0^2 + \frac{3 }{x_s} z_0 -1 = 0.
\end{equation}
The solution to Eq.~\eqref{cubic} depends on the value of $r_s$. For $r_s < a/\sqrt[3]{2} \approx 0.4872$, there are 3 real roots but only 1 is positive and for larger $r_s$ there is only 1 real root. Thus, 
\begin{equation}
\label{zSoln}
z_0 = \begin{cases}
Q^{1/2} \left[ \; 2 \textrm{cos} \left( \frac{\theta + \pi}{3} \right) + \sqrt{3}\; \right] & r_s < a/\sqrt[3]{2} \\
A + Q/A - \sqrt{3Q}& r_s > a/\sqrt[3]{2},
\end{cases}
\end{equation}
where $\theta = \textrm{cos}^{-1} (\sqrt{2 x_s^3})$, $Q = 1/2 x_s$, and $A = [ 1/2 + (1/4 - Q^3)^{1/2} ]^{1/3}$. The numerical solution to the TF blue electron model is detailed in Appendix D.

\begin{figure}[htb]
\begin{center}
\includegraphics[angle=0,width=8.5cm]{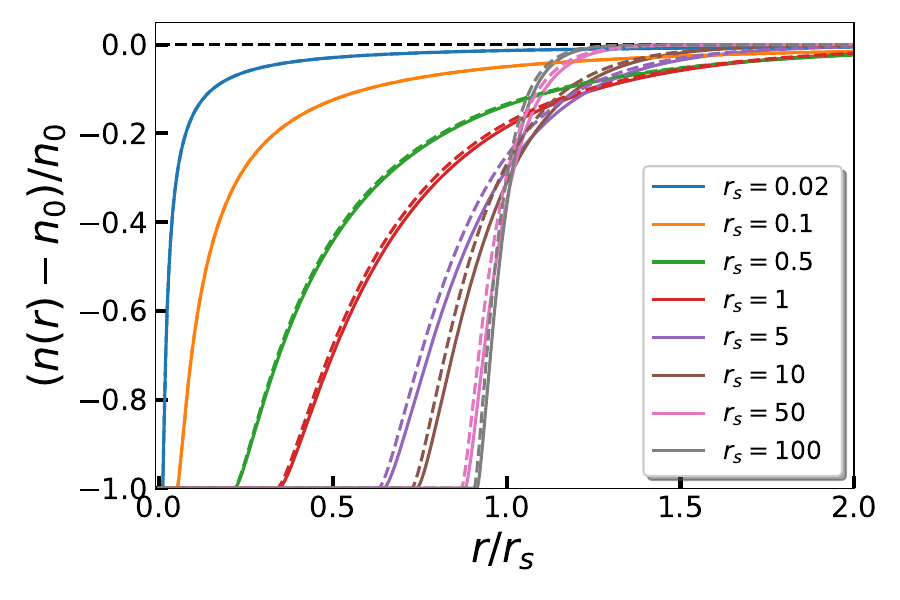}
\end{center}
\caption{Normalized hole density vs $r/r_s$ for a range of densities. Solid lines are the numerical solutions and the dashed lines are for the analytic solutions.}
\label{f:TF_f6a}
\end{figure}

\begin{figure}[htb]
\begin{center}
\includegraphics[angle=0,width=8.5cm]{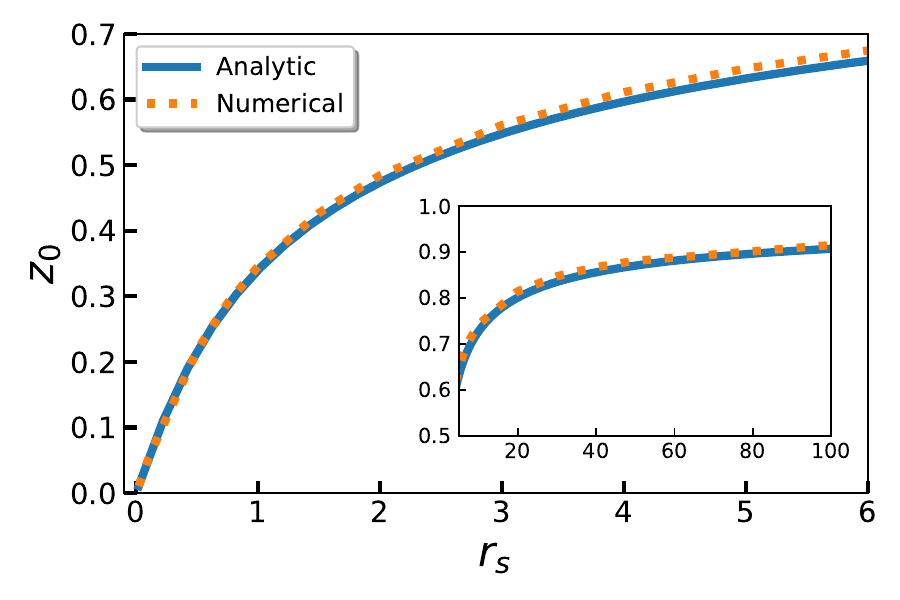}
\end{center}
\caption{Value of $z_0$ vs $r_s$. The solid line is derived from Eq.~\eqref{cubic} and the dotted line is from the numerical calculations}.
\label{f:TF_f7}
\end{figure}

We performed calculations for $r_s$ values ranging from $r_s = 0.02$ to $r_s = 100$. Plots of the normalized hole densities for these $r_s$ values are shown in Fig.~\ref{f:TF_f6a}. As can be seen in this figure, the normalized hole density tends toward a step function at $r = r_s$ in the low density limit and the location of $z_0$ moves farther out. In Fig.~\ref{f:TF_f7} we plot the value of $z_0$ as a function of $r_s$ as determined from the numerical calculations, as well as the analytic approximation of Eq.~\eqref{zSoln}. The analytic solution for $z_0$ is surprisingly accurate with a maximum error of 2.3\% around $r_s = 5.0$. A table of values for the numerical and analytical results for $z_0$ can be found in Table~\ref{Z0_Values}.

\begin{figure}[htb]
\begin{center}
\includegraphics[angle=0,width=8.5cm]{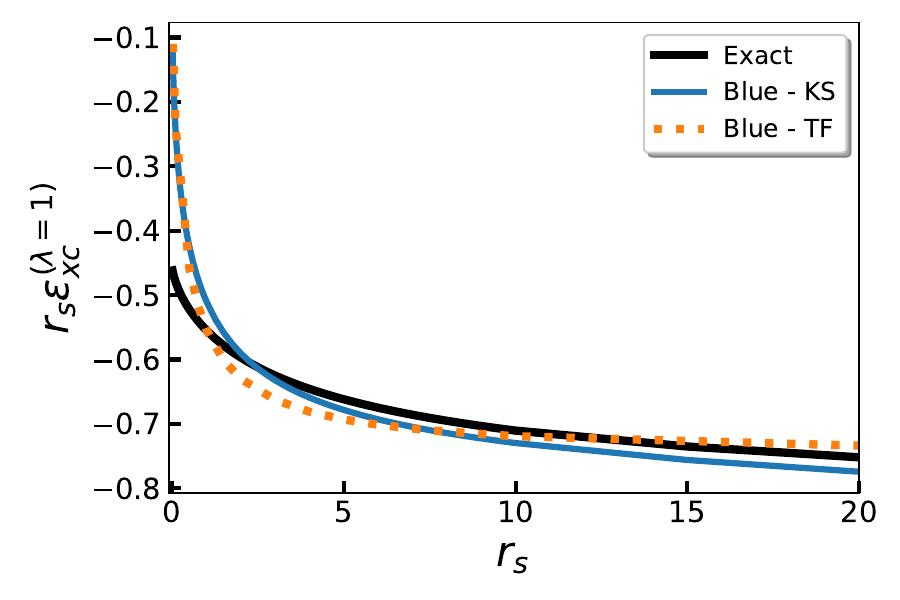}
\end{center}
\caption{$r_s \varepsilon_{\xc}^{(\lambda=1)}$ vs $r_s$ for the TF Blue Electron model , the KS Blue Electron model, and the exact result. }
\label{f:TF_f8}
\end{figure}

\begin{figure}[htb]
\begin{center}
\includegraphics[angle=0,width=8.5cm]{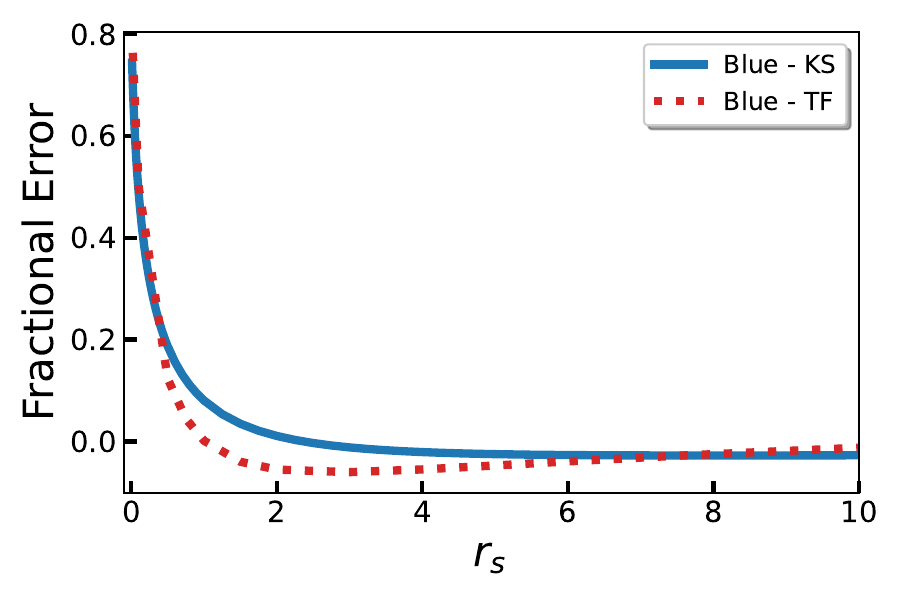}
\end{center}
\caption{Fractional error for the TF model), KS model, and the exact result. The comparisons are with the non-coupling-constant averaged energies}.
\label{f:TF_f9}
\end{figure}

\begin{figure}[htb]
\begin{center}
\includegraphics[angle=0,width=8.5cm]{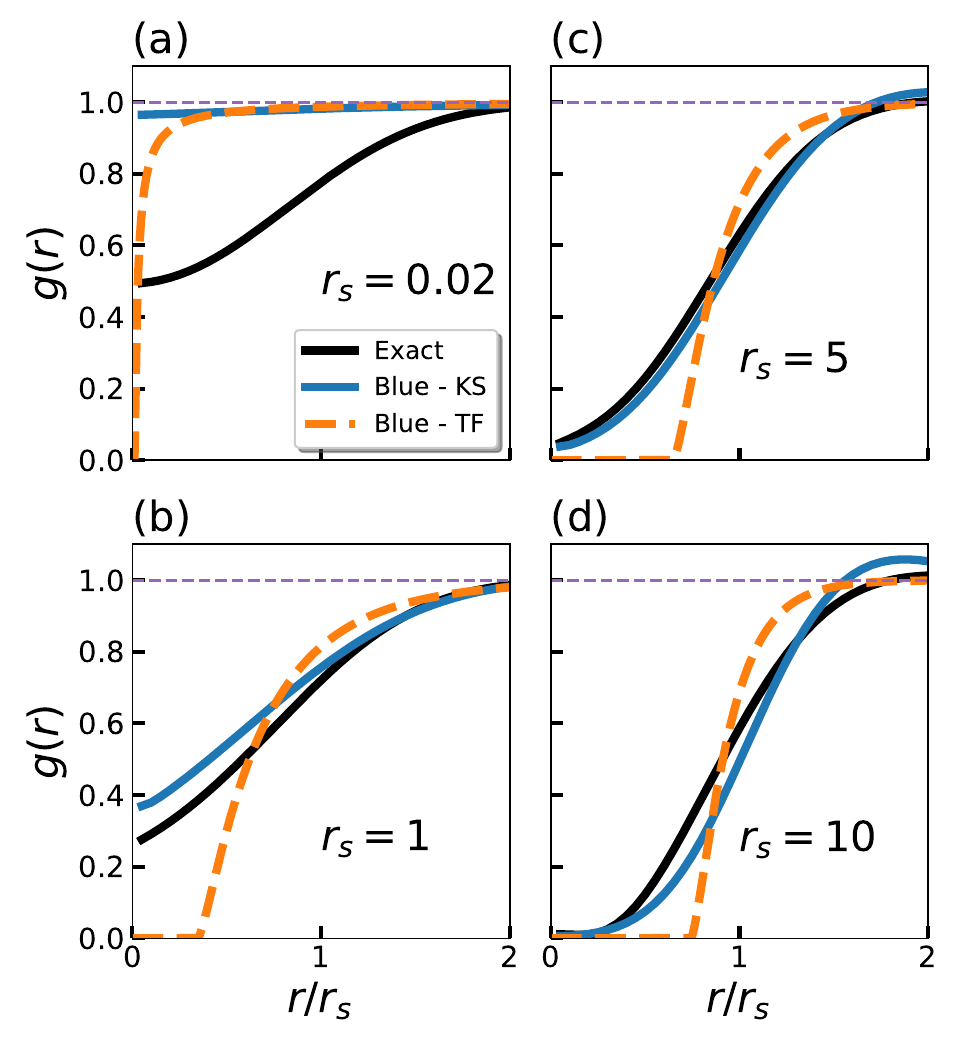}
\end{center}
\caption{PDF vs $r/r_s$ for (a) $r_s = 0.02$, (b) $r_s = 1.0$, (c) $r_s = 5.0$, (d) $r_s = 10.0$  The solid black line is the exact result, the solid blue line is the CP-DFT with $\Delta \tilde{v}= 1/r$, and the dashed orange line is the Thomas-Fermi model.}
\label{f:TF_f15}
\end{figure}

\begin{table}[ht]
\begin{center}
\renewcommand{\arraystretch}{1.4}
\begin{tabular}{|c|c|c|c|} 
\hline
$r_s$ & $z_0$ & $z_0^{A}$ & $ (z_0^A - z_0)/|z_0| $ \\
\hline
0.02	& 0.0108	& 0.0108	& -0.0000\\
\hline
0.1	 & 0.0529	 & 0.0529	 & -0.0004\\
\hline
0.5	& 0.2168	& 0.2186	& -0.0081 \\
\hline
 1.0	& 0.3404	& 0.3460	& -0.0161\\
\hline
5.0	& 0.6322	& 0.6472	& -0.0233 \\
\hline
10.0	& 0.7272	& 0.7425	& -0.0205 \\
\hline
50.0	& 0.8704	& 0.8809	& -0.0119 \\
\hline 
100.  & 0.9071	 & 0.9153	& -0.0089 \\
\hline
\end{tabular}
\caption{Values of $z_0$ as a function of $r_s$ exactly and analytic calculations along with the relative error between them.}
\label{Z0_Values}
\end{center}
\end{table}
We can calculate the potential XC energy from the CP hole density, $n_{\xc} (r) = n(r) - n_0$ using Eq.~\eqref{eq:exc}.
In Fig.~\ref{f:TF_f8} we plot potential energy $r_s \varepsilon_{\xc}^{(\lambda=1)}$ and compare with the the exact result as well as the KS blue electron result using CP-DFT with $\Delta \tilde{v}(r) = 1/r$. We also plot the fractional error in Fig.~\ref{f:TF_f9}. For mid- to low-densities ($r_s \geq 1.0$) the results are surprisingly good, with errors less than 10\%. It is instructive at this point to plot the PDF vs $r/r_s$ for different values of $r_s$ and compare with the exact result. This comparison is shown in Fig.~\ref{f:TF_f15} for $r_s = 0.02, 1.0, 5.0, 10.0$. The results are poor at $r_s = 0.02$ but improve with decreasing density. Interestingly enough, although the shape of the hole is inaccurate at $r_s = 1.0$, the integrated value of the hole energy is quite good when compared to the exact result.

\sec{Conclusions}
We have shown how to implement the blue-electron concept for a uniform gas at low
temperature.   The simple classical idea of an impurity capture the basic physics for
moderate to low densities, and its accuracy is improved by a local density approximation for
the CP potential that respects the electron-electron cusp condition at short distances.

The blue-electron idea naturally fails at high densities, where exchange effects dominate.  We have shown how this failure can be overcome by applying the idea for anti-parallel spin
alone, while using just the exchange hole for parallel spin.  Our procedure still avoids any many-body or QMC input, and provides useful accuracy for all densities, including the high-density limit.  Moreover, it provides formal justification for the large Gaussian repulsion used when the entire hole is simulated for high-densities, and its smooth turning off.  Errors no greater than 10\% in the XC energy density are then available.

As the method is based on classical DFT ideas, we expect errors to lessen with increasing temperature, and overall, they do \cite{MPPE20b}. But we have also given results for TF blue electron calculations, as higher-temperature simulations often must use TF in place of KS because
of convergence issues with the KS scheme. Here we present our analysis of the TF equation just at zero temperature.A simple analytic approximation is remarkably accurate relative to a fully converged numerical solution.We use the numerical solution to compare with KS blue electron calculations at zero temperature.   Again, we expect the error introduced by the TF approximation to reduce with increasing temperature.

Even at low temperature, where convergence of KS is not an issue, the TF blue electron might prove a pragmatic alternative to full CP-DFT calculations, as these require a KS calculation
at every point in the system.

Regardless of the accuracy of these specific calculations, the principle that one could
calculate XC holes and their energies exactly via the densities of many KS calculations
remains valid.   Moreover, the CP approach provides a useful bypass of the need to find more accurate XC functionals, albeit at higher computational cost.   It remains to be seen whether CP-DFT can evolve into a practical alternative to existing DFT calculations, especially in situations where standard DFT fails.

\begin{acknowledgments}
D.P. and K.B. supported by DOE DE-FG02-08ER46496,
R.J.M. by a University of California
President's Postdoctoral Fellowship and the National Science Foundation (CHE 1856165). 
\end{acknowledgments}

\appendix
\section{CP-DFT details}

Here we provide further details of the CP-DFT and our method of numerical solution. In addition to the Hartree, Eq.~\eqref{ve} and  CP potentials Eq.~\eqref{vBKS} used in the KS equations, Eq.~\eqref{KSBlueNS}, we have the exchange interaction in the LDA,
\begin{equation}
\label{vx}
v\x^{\sss LDA}[n](r) = \left( \frac{3}{2\pi} \right)^{2/3} \frac{1}{r_s[n(r)]},
\end{equation}
and,  when used,  the correlation potential, $v\c[n](r)$ is from PW92.
The positive, compensating background potential is
\begin{equation}
\label{vn}
v_{\sss n}(\mathbf{r}) = -\int d^3r' \frac{n_0}{|\mathbf{r-r'}|} = -4 \pi \bar{n} \left(  \frac{R^2 }{2}- \frac{r^2}{6}  \right).
\end{equation}

Our system is spherically symmetric, hence we can write the wave functions in terms of spherical harmonics and a radial wave function,
\begin{equation}
\phi_i(\mathbf{r}) = R_{nl}(r) Y_{lm}(\theta,\phi) = \frac{u_{nl}(r)}{r} Y_{lm}(\theta,\phi),
\end{equation}
with quantum numbers $n,l,m$ and we have expressed the radial wave function as $u(r)/r$.  The radial KS equation is
\begin{equation}
\label{eq:KSR}
\left[ -\frac{1}{2} \frac{d^2}{dr^2} + v\s(r) + \frac{l(l+1)}{2r^2} \right] u_{nl}(r) = \varepsilon_{nl}u_{nl}(r).
\end{equation}
The density is 
\begin{equation}
\label{eq:NSDensity}
n(r) = 2 \sum_{n,l} \frac{2l+1}{4\pi} \frac{u_{nl}^2(r)}{r^2} f((\varepsilon_{nl}-\mu)/k_BT) ,
\end{equation}
where the factor of $2$ accounts for spin and $f((\varepsilon_{nl}-\mu)/k_BT)$ is the Fermi-Dirac distribution function used for Fermi \textit{smearing} of the energy states,
\begin{equation}
f(x) = (1 + e^x)^{-1}.
\end{equation}
Here, $\mu$ is the chemical potential chosen so that
\begin{equation}
\label{eq:mu}
N = 2\sum_{n,l} (2l+1)f((\varepsilon_{nl}-\mu)/k_BT),
\end{equation}
where $N$ is the number  electrons and $k_BT$ is Boltzmann's constant times the temperature.  Smearing reduces convergence issues. All calculations use  $\phi = 0.05$ defined as
\begin{equation}
\label{eq_TR}
\phi = \frac{k_bT}{\varepsilon_F^{ni}},
\end{equation}
where $\varepsilon_F^{ni}$ is the Fermi energy for the non-interacting electron gas
\begin{equation}
\varepsilon_F^{ni} = \frac{1}{2} \left( \frac{9\pi}{4} \right)^{2/3} r_s^{-2}.
\end{equation}

The set of equations describing the CP-DFT are coupled and  thus have to be solved self-consistently. We do this using an iterative scheme wherein first, an initial guess is made for the  density, $n(r)$. Then, the KS equations are solved using that initial guess and a new density, $n^{*}(r)$ evaluated from Eq.~\eqref{eq:NSDensity}.  This density is then mixed in with the previous density via linear mixing:
%primitive mixing:
\begin{equation}
n^{k+1}(r) = \alpha n^{*}(r) + (1-\alpha) n^k (r),
\end{equation}
where $k$ represents the $k^{th}$ iteration and $\alpha <1$.
This process is repeated until a suitable convergence to be described below is obtained. We use a shooting method to solve for the wave function that satisfies Eq.~\eqref{eq:KSR} for a given energy, $\varepsilon_{nl}$. A $4^{th}$ order Runge-Kutta scheme is used to integrate Eq.~\eqref{eq:KSR}  forward from $r = r_0$ to $r = R$ with typical values $r_0 = 0.05 r_s$ and $R = 8r_s$, respectively. Following \cite{Nogueira2003} the boundary condition used at $r = r_0$  is
\begin{equation}\label{eq:BndryL}
\begin{aligned}
u_{nl}(r=r_0) =& Ar_0^{l+1} \\
\left. \frac{du}{dr} \right|_{r = r_0} =& A(l+1)r_0^l,
\end{aligned}
\end{equation}
where $A$ is some constant.  For the boundary condition at $r=R$, we require  $\left. \partial n(r) / \partial r \right|_{r=R} = 0$.
To ensure this we require
\begin{equation}
\label{eq:noflux2}
\left. \left( \frac{\partial u(r)}{\partial r} - \frac{u(r)}{r} \right) \right|_{r=R} = 0.
\end{equation}
Solving the KS equations for a specific $l$  involves varying the energy $\varepsilon$ in small increments, integrating Eq.~\eqref{eq:KSR} for each value of the energy and looking for regions where Eq.~\eqref{eq:noflux2} changes sign. A root-finding scheme is then used to precisely locate the energy eigenvalue.

We use two criteria for convergence of the iterative solution scheme.  The first is that the difference of the sum over all states of the energy eigenvalues between the new state before mixing and the previous iteration relative to the new state must be less than $5 \times 10^{-5}$. 
%$5 x 10^{-5}$. 
The second is that the 2-norm of the density difference between successive iterations normalized by the 2-norm of the density is less than $.001$. Both criteria must be satisfied for convergence.

The procedure for a blue electron calculation is then to solve both the $N$ electron system with $\Delta \tilde{v} = 0$ and the $N-1$ electron system with $\Delta \tilde{v}$ as in Eq.~\eqref{vBKS} for a given system density. With the densities obtained from these two calculations, $n(r)$ and $\tilde{n}(r)$, we calculate the XC hole density as
\begin{equation}
\label{eq:hole}
n\xc^{\lambda=1}(r) = \tilde{n}(r) - n(r)
\end{equation}
and from the hole density obtain the PDF
\begin{equation}
\label{eq:pdf}
g(r) = \frac{n\xc^{\lambda=1}(r)}{\bar{n}}+1.
\end{equation}
The XC energy can be calculated as
\begin{equation}
\label{eq:exc}
\varepsilon\xc^{(\lambda)} = 2 \pi \int_0^R r \; n\xc^{(\lambda)}(r)  dr.
\end{equation}

\section{Spin CP-DFT details}
The spin CP-DFT considers the up-spin density fixed by Eq.~\eqref{eq:upspin} and solves for the down spin density, $n_{\uparrow}(r \downarrow)$.
Thus we need an expression for the like-spin PDF.
An analytic fit to the like-spin PDF can be obtained from Eq. 48 of PW92
\begin{equation}
\label{eq:gUU}
g_{\uparrow \uparrow}(r_s,\zeta,k_Fr) = g\left[r_s, 1, (1 + \zeta)^{1/3} k_Fr/2^{1/3} \right],
\end{equation}
where $r_s = (3/4\pi \bar{n})^{1/3}$ is the Wigner-Seitz radius, $\bar{n} = \bar{n}_\uparrow + \bar{n}_\downarrow$ , $k_F = (3 \pi^2 \bar{n} )^{1/3}$ is the Fermi wave vector, $\zeta = (\bar{n}_\uparrow - \bar{n}_\downarrow)/(\bar{n}_\uparrow + \bar{n}_\downarrow)$ is the spin polarization.  In our calculations, we include only the exchange part of the like-spin PDF
\begin{equation}
g_{\sss X}(\zeta,k_Fr) = 1 + \frac{1}{2} \sum_{\sigma=-1}^{+1} (1+\sigma \zeta)^{2} J [ (1+\sigma \zeta)^{1/3}k_Fr ],
\label{gx}
\end{equation}
where
\begin{equation}
J(y) = -\frac{9}{2} \frac{j_1(y)}{y}
\end{equation}
and 
$j_1$ is the spherical Bessel function of the first kind.
Similar to the neutral spin case, we can write the KS orbitals for $n_\downarrow(r)$ in terms of spherical harmonics and a radial function. This leads to a KS equation for $n_\downarrow(r)$ similar to Eq.~\eqref{eq:KSR} but without the 2 for spin degeneracy. 

The KS potential is as before but now the exchange potential is:
\begin{multline}
v^{LDA}_{\sss X,\sigma}[n,\zeta](r) = \frac{4}{3} \varepsilon_x(n,\zeta) \\
- \frac{2}{3} \left( \zeta - \mathrm{sign}(\sigma) \right) \varepsilon_{\sss X}^{\sss{unpol}}(n) \left\{ (1+\zeta)^{1/3} - (1-\zeta)^{1/3} \right\}.
\end{multline}
% \begin{multline}
% v\x^\sigma(n,\zeta) = \frac{4}{3} \varepsilon_x(n,\zeta) \\
% - \frac{2}{3} \left( \zeta - \mathrm{sign}(\sigma) \right) \varepsilon\x^{\sss{unpol}}(n) \left\{ (1+\zeta)^{1/3} - (1-\zeta)^{1/3} \right\}
% \end{multline}
Here, $\text{sign}(\sigma) = \pm 1$ as $\sigma = \uparrow$ or $\sigma = \downarrow$. The unpolarized exchange energy is
\begin{equation}
\varepsilon\x^{\sss{unpol}} (n)  = -\frac{3}{4} \left( \frac{3}{\pi} \right)^{1/3} n^{1/3},
\end{equation}
and the polarized exchange energy is
\begin{equation}
\varepsilon_{\sss X}(n,\zeta) = \varepsilon_{\sss X}^{\sss{unpol}} (n)\left\{ \frac{(1+\zeta)^{4/3} + (1-\zeta)^{4/3}}{2}\right\}.
\end{equation}
Note that in the above equations, $\zeta = \zeta(r)$ is  the local spin polarization. The correlation potential, when used, is from PW92 and the CP potential is as before.

The $n_\uparrow(r)$ is kept fixed and the KS equations are solved for the down spin density in a manner analogous to that of the CP-DFT case. Once the down spin densities are obtained, the PDF can be calculated as before using Eq.~\eqref{eq:hole} and Eq.~\eqref{eq:pdf}.
Additionally we can calculate the antiparallel PDF from
\begin{equation}
g_{\uparrow \downarrow}(r) = \frac{n_{\sss XC,\downarrow}(r)}{\bar{n}_{\downarrow}}  + 1,
\end{equation}
with
\begin{equation}
n_{\sss XC,\downarrow}^{(\lambda=1)}(r) = \tilde{n}_{\uparrow}(r \downarrow) - n(r \downarrow).
\end{equation}

\section{Determination of parameters for the Gaussian potential}
The parameters for this added potential were obtained by varying values of $A$ and $\sigma$ in a grid search and evaluating two metrics. The first is the relative difference with PW92 for the on-top hole density at $r=0$, i.e.,
\begin{equation}
\label{eq:MeaureA}
M_A = \frac{(n_{xc}(r\rightarrow 0)-n_{xc}^{PW}(r\rightarrow 0))}{|n_{xc}^{PW}(r\rightarrow 0)|}
\end{equation}
and the second is the relative difference with the PW92 exchange-correlation energy,
\begin{equation}
\label{eq:MeasureB}
M_B = \frac{(\varepsilon_{xc}-\varepsilon_{xc}^{PW})}{|\varepsilon_{xc}^{PW}|}.
\end{equation}

\begin{figure}[htb]
\begin{center}
\includegraphics[angle=0,width=8.5cm]{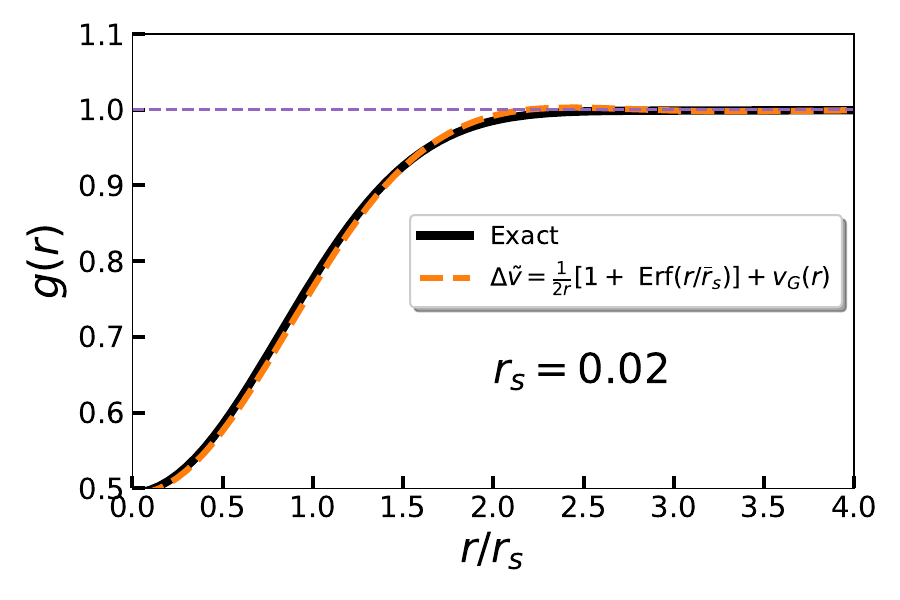}
\end{center}
\caption{Pair distribution function, $g(r)$ plotted vs $r/r_s$ for $r_s = 0.02$. The solid black line is the exact PDF and the dashed orange line is for the  modified CP potential plus the added Gaussian of Eq.~\eqref{eq:Gaussian}.}
\label{f:f_C1}
\end{figure}

The zero-contours from the grid search of each of these metrics was calculated and the values for $A$ and $\sigma$  determined from the intersection of these zero-contours. The results at $r_s = 0.02$ were $A = 2322.2065$ and $\sigma = 0.01145$. Using these values in Eq.~\eqref{eq:Gaussian} matches the PW92 PDF for $r_s = 0.02$ very well as can be seen in Fig.~\ref{f:f_C1}. 

The behavior of $A$ and $\sigma$ as determined by additional grid searches appeared to have power-law dependencies, at least in the region of high densities (low $r_s$).
However, this held true only for a very narrow range of $r_s$ values and quickly departed from this behavior as $r_s$ increased. We decided that the parameters of the Gaussian potential would have this power-law behavior but would be multiplied by a switching function such that by the time $r_s = 2.5$ was reached, the Gaussian potential parameters vanished. Thus, we used the following
\begin{equation}
\label{eq:powerS}
\begin{aligned}
A(r_s) =& A_0  f_s(r_s,r_R,b_A)/r_s^2 \\
\gamma (r_s) =& \gamma_0  f_s(r_s,r_R,b_G)/r_s,
\end{aligned}
\end{equation}
where
\begin{equation}
\sigma(r_s) = \frac{1}{\gamma(r_s)}
\end{equation}
and the switching function, $f_s$ is given by \cite{PCK_2010}:
\begin{equation}
\label{eq:switch}
f_s(r,r_R,b) = \begin{cases}
\frac{1}{1-e^{-b}} \left( e^{-b \left(\frac{r}{r_R} \right)} -e^{-b} \right) & 0 \leq r \leq r_R \\
0 & r_R \leq r.
\end{cases}
\end{equation} 
Here, $r_R$ denotes the range over which the switching function goes from one to zero and $b$ is a parameter controlling the shape of the switch. In the limit of $b \rightarrow 0$, the switch is a linear function. The utility of this function is the wide range of switching behaviors that it can display, from rapid decrease and subsequent slower decline to slow decline followed by rapid decrease and as mentioned above, a simple linear decline.

The procedure we applied to determine the parameters was as follows. We would choose a set of parameters for the switching functions, i.e., $b_A, \; b_\gamma$  (Note that we use the same $r_R$ parameter for both $A$ and $\gamma$ but allow for different $b_{A(\gamma)}$ values.) The value for $r_R$ was  set to $r_s = 2.5$. The parameters $A_0$ and $\gamma_0$ are determined from the values obtained from the grid search at $r_s = 0.02$,
\begin{equation}
\begin{aligned}
A_0 =& \frac{A(r_s = 0.02) (0.02)^2} {f_s(0.02,r_R,b_A)} \\
\gamma_0 =& \frac{ \gamma(r_s = 0.02) (0.02) }{f_s(0.02,r_R,b_\gamma)}.
\end{aligned}
\end{equation}
Then, $b_A $ and $b_\gamma$ are varied until there is a smooth to the eye transition from the $r_s = 0.02$ value to the $r_s = 2.5$ value when plotting $r_s \varepsilon_{xc}$ vs $r_s$. The values of $r_R = 2.5, \; b_A = 3.0, \; b_\gamma = -3.0$ appeared to work best. 
%\label{page:end}
\section{Numerical method for the Thomas-Fermi blue electron model}
\begin{figure}[htb]
\begin{center}
\includegraphics[angle=0,width=8.5cm]{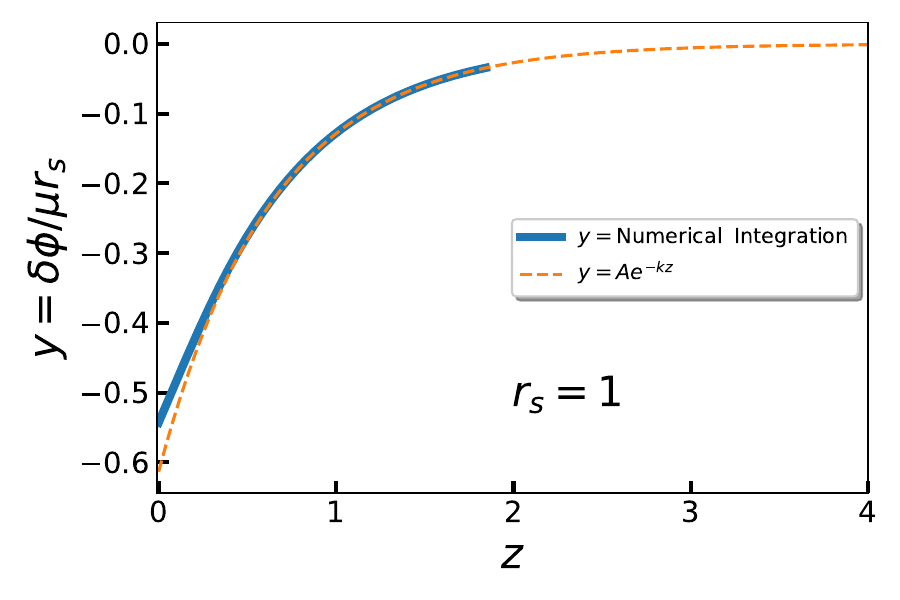}
\end{center}
\caption{The solution to Eq.~\eqref{diffY} for $r_s = 1.0$. The solid line is the numerical solution and the dashed line is the large $z$ solution with the value of $A_0$ given by Eq.~\eqref{A0Match}.}
\label{f:TF_f2}
\end{figure}

\begin{figure}[htb]
\begin{center}
\includegraphics[angle=0,width=8.5cm]{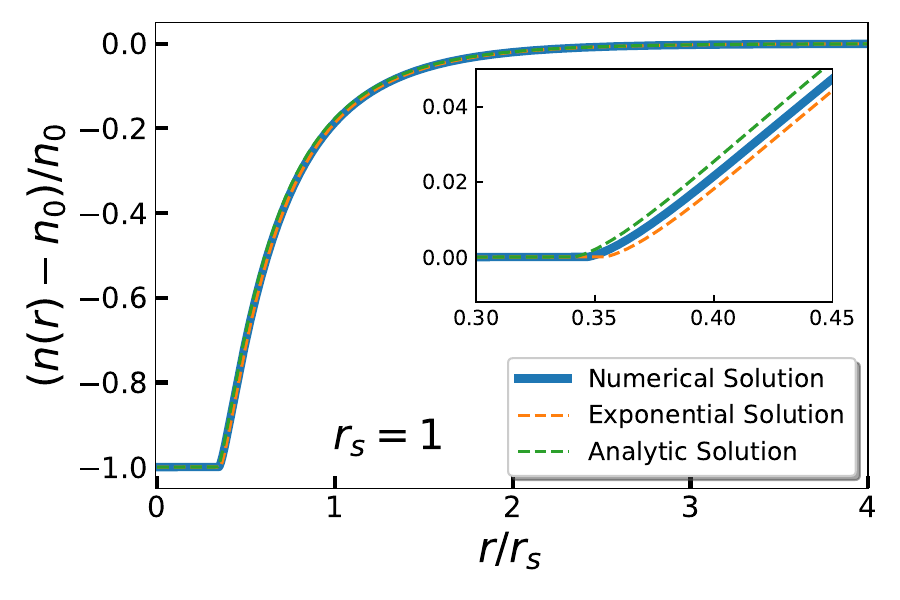}
\end{center}
\caption{Normalized hole density vs $r/r_s$ for $r_s =1.0$ Solid line is the result from the numerical calculation whereas the orange dashed line is from the exponential solution with $A_0$ determined from Eq.~\eqref{A0Match} while the dashed green line is the analytic solution with $A_0$ determined by the boundary conditions. The inset shows the region for $r_s \lessapprox 0.5$}
\label{f:TF_14}
\end{figure}

As discussed above, we only expect Eq.~\eqref{cubic} to be valid at high densities. Nonetheless, it can used to get an initial approximation for $z_0$ at all densities. Using the solution to Eq.~\eqref{cubic} as an initial guess for $z_0$, we can numerically integrate Eq.~\eqref{diffY} starting from $z = 0$ with the boundary conditions Eq.~\eqref{BoundY}. We continue the numerical integration until certain criteria are reached. First, it may occur that $y(z) > 0$. In this case, our initial guess for $z_0$ is too low and needs to be increased. Second, it may occur that $d y(z)/dz < 0$. In this case, the solution is tending away from zero which is where the solution must go as $z \rightarrow \infty$. Here, $z_0$ must be decreased. Lastly, we may find that $d\, ln(y)/dz = -k$ to within some designated tolerance. This means that we have effectively reached the large $z$ region of Eq.~\eqref{diffY} for the particular value of $r_s$ being used and we can now switch to the analytic result to obtain the solution for larger $z$ values. Once Eq.~\eqref{diffY} has been integrated to a point, $z = z_{ldm}$ where the log derivative matches $-k$ to the specified tolerance, the value of the constant, $A_0$ can be determined from 
\begin{equation}
\label{A0Match}
A_0 = y(z_{ldm})e^{kz_{ldm}}.
\end{equation}
We used a $4^{th}$ order Runge-Kutta scheme with a step-size of $10^{-5}$ to integrate Eq.~\eqref{diffY}. Initial work showed that the value of $z_0$ obtained from Eq.~\eqref{cubic} is slightly below that of the actual value and so we created a bracket with the lower guess, $z_0^L$ from Eq.~\eqref{cubic} and the upper guess being $z_0^R = z_0 + 0.1$. We then solve Eq.~\eqref{diffY} using $z_0 = (z_0^L+z_0^R)/2$ until one of the above stopping criteria are reached. If the log derivative matches to that of $-k$ with a tolerance of $1.0x10^{-8}$, then we have found $z_0$. If, one of the other criteria is reached, we adjust either the lower or upper guess and repeat the process until the log derivative is matched. This iterative process required some 10-50 iterations depending on the density.
%An example of this iterative process is shown in Fig.~\ref{f:TF_f1} for $r_s = 1.0$.
%\begin{figure}[htb]
%\begin{center}
%\includegraphics[angle=0,width=8.5cm]{TF_f1.pdf}
%\end{center}
%\caption{Iterated value of $z_0$ vs iteration for $r_s =1.0$}.
%\label{f:TF_f1}
%\end{figure}
The numerical solution to Eq.~\eqref{diffY} for $r_s = 1.0$  with the value of $z_0 = 0.3460$ found by the iterative process is shown in Fig.~\ref{f:TF_f2} along with the exponential solution with $A_0$ obtained from Eq.~\eqref{A0Match}
The normalized hole density, $(n(r)-n_0)/n_0$ for the case of $r_s = 1.0$ is shown in Fig.~\ref{f:TF_14} along with the exponential solution with $A_0$ obtained from Eq.~\eqref{A0Match} and an analytic solution where $A_0$ is obtained from the boundary conditions, Eq.~\eqref{BoundY} using the $z_0$ value obtained from Eq.~\eqref{cubic}.  A closer look at the region for $r/r_s < 0.5$ is shown in the inset where we can see the differences between the numerical solution and the exponential and purely analytic solutions.
As a last check on the calculation, we can calculate the integral of the hole density for the different solutions and it should integrate to $-1$. The numerical and exponential solutions satisfies this condition to about 0.04\% and 0.40\% respectively whereas the purely analytic solution is off by approximately 5\%. We note that for matching the log derivatives, a tolerance of $10^{-8}$ is necessary to have the hole density integrate to $-1$. At lower densities, a lower tolerance of $10^{-6}$ works just fine. 

%\begin{figure}[htb]
%\begin{center}
%\includegraphics[angle=0,width=8.5cm]{TF_f5.pdf}
%\end{center}
%\caption{Cumulative integral of the hole density vs $r/r_s$ for $r_s =1.0$ for the different solutions}.
%\label{f:TF_f5}
%\end{figure}

\newpage
\bibliographystyle{apsrev4-2}
\bibliography{Slimbib}

\end{document}